\documentclass[twocolumn,prb,preprintnumbers,amsmath,amssymb]{revtex4}

\usepackage{graphicx}
\usepackage{dcolumn}
\usepackage{bm}
\usepackage{hyperref}
\usepackage{amsmath}
\usepackage{epsfig}

\begin{document}
\title{Dispersion of waves in relativistic plasmas with isotropic particle distributions}

\author{Roman V. Shcherbakov}
\email{rshcherbakov@cfa.harvard.edu} \homepage{http://www.cfa.harvard.edu/~rshcherb/}
\affiliation{Harvard-Smithsonian Center for Astrophysics, 60 Garden Street, Cambridge, MA 02138}
\date{\today}

\begin{abstract}
The dispersion laws of Langmuir and transverse waves are calculated in the relativistic
non-magnetized formalism for several isotropic particle distributions: thermal, power-law,
relativistic Lorentzian $\kappa,$ and hybrid $\beta$. For Langmuir waves the parameters of
superluminal undamped, subluminal damped principal and higher modes are determined for a range of
distribution parameters. The undamped and principal damped modes are found to match smoothly.
Principal damped and second damped modes are found not to match smoothly. The presence of maximum
wavenumber is discovered above that no longitudinal modes formally exist. The higher damped modes
are discovered to be qualitatively different for thermal and certain non-thermal distributions.
Consistently with the known results, the Landau damping is calculated to be stronger for
non-thermal power-law-like distributions. The dispersion law is obtained for the single undamped
transverse mode. The analytic results for the simplest distributions are provided.
\end{abstract}
\maketitle

\section{Introduction}\label{sec_intro}
The non-thermal distributions of electrons are as important as thermal for astrophysical plasmas.
Shocked and turbulent medium is likely to accelerate the electrons into the power-laws. Gamma-Ray
bursts\citep{panaitescu} (GRBs), jets from compact sources\citep{wardle}, low luminosity active
galactic nuclei\citep{yuan} (LLAGNs) show evidence of the non-thermal relativistic distributions.
The electrons are non-thermal and mildly relativistic near Earth\citep{xiao_earth} and in solar
corona\citep{maksimovic}.

The properties of EM waves propagating through such medium are worth knowing even in a
non-magnetized case. EM waves can be generated in turbulence, propagate some distance and dump via
the Landau damping. Thus energy is redistributed. Whereas the realistic turbulence is non-linear,
only the linear waves are considered in this work. No back-reaction of waves on the electron
distributions is assumed.

The reasonable relativistic electron distributions are thermal (\ref{distth}), power-law
(\ref{distpower_law}), Lorentzian $\kappa$ (\ref{distkappa}) and hybrid $\beta$ (\ref{distbeta}).
They are taken to be normalized to unity when integrated over momenta. I take all the quantities in
the paper to be dimensionless for the sake of clarity and brevity. Temperature $T$ of thermal
distribution is evaluated in the units of $m c^2/k_B$ for the particles with mass $m.$ Here $c$ is
the speed of light and $k_B$ is Boltzmann constant. The inverse temperature
\begin{equation}
\rho=\frac1T
\end{equation}
is sometimes denoted as $\mu$ in the literature. Momenta of particles $p$ is measured in $m c.$ The
speed of light $c$ is set to unity. Thus
\begin{equation}
\gamma=\sqrt{1+p^2}
\end{equation} is the particles dimensionless energy.
The non-relativistic plasma frequency in CGS units (for particles with charge $q$)
\begin{equation}
\omega_p=\sqrt{\frac{4\pi n q^2}{m}}
\end{equation} is employed to normalize the absolute values of a wavenumber $k$ and frequency
$\omega$ as\citep{bergman}
\begin{equation}
\Omega=\frac{\omega}{\omega_p}, \qquad {\bf K}=\frac{{\bf k}}{\omega_p}.
\end{equation} The scalar $K$ is the absolute value of the vector ${\bf K}.$

The general characteristics of relativistic plasma are the following. Transverse waves have the
phase velocity greater than the speed of light $\Omega/K>1$ and so are undamped. They only have the
single mode. The longitudinal waves exhibit the wider range of phenomena\citep{schlick_kin,
bergman}. The phase velocity goes from $>1$ to $<1$ as $K$ increases, the undamped mode at small
$K$ matches the principal damped mode that exists at higher $K$ up to
$K_{max}.$\citep{schlick_kmax} Besides the principal mode, the finite set of higher damped modes
exists. The present paper elaborates on all these modes for various isotropic particle
distribution. Either a mixture of electrons and positrons with same distributions or immobile ions
are considered, so that the ion sound does not appear.

The paper is organized as follows. The formalism of linear plasma dispersion is reviewed in \S
\ref{sec_linear}. The results of numeric evaluations are presented in \S \ref{sec_numeric}. Some
analytic formulas can be found in \S \ref{sec_analytic}.
I conclude with the discussion and the summary in \S \ref{sec_discussion}.

\section{Linear dispersion laws}\label{sec_linear}
Waves propagating in plasma feel the plasma response, which can be characterized by the
permittivity tensor $\varepsilon({\bf K},\Omega)$ in the linear regime.
Permittivity tensor\citep{imre,bergman}
\begin{equation}
\varepsilon({\bf K},\Omega)=1+\sum_{species}\frac1{\Omega^2}\int\frac1p\frac{\partial f}{\partial
p}\frac{\bf p p}{\gamma -{\bf K\cdot p}/\Omega}d^3p
\end{equation} leads\citep{bergman,melrosec_q}  after the integration over the polar angles to longitudinal
permittivity
\begin{eqnarray}
\varepsilon_L&=&\frac{2\pi^2 i \sigma \Omega}{K^3}\int^\infty_{\gamma_0}\frac{\partial
f}{\partial \gamma}\gamma^2 d\gamma +1\\
&-&\frac{2\pi}{\Omega K}\int^\infty_0p^2\frac{\partial f}{\partial p}\Bigg\{\frac{2\gamma \Omega}{K
p}+\frac{\gamma^2\Omega^2}{K^2 p^2}\ln\left[\frac{\gamma\Omega-K p}{\gamma\Omega+K
p}\right]\Bigg\}dp \nonumber
\end{eqnarray} and transverse permittivity
\begin{eqnarray}
\varepsilon_T=1&+&\frac{\pi^2 i\sigma}{\Omega K}\int^\infty_{\gamma_0}\frac{\partial f}{\partial
\gamma}\bigg(1-\frac{\gamma^2}{\gamma_0^2}\bigg)d\gamma+\\\frac{\pi}{\Omega K}\int^\infty_0
p^2\frac{\partial f}{\partial p}\Bigg\{\frac{2\gamma \Omega}{K
p}&+&\left(\frac{\gamma^2\Omega^2}{K^2 p^2} - 1\right)\ln\left[\frac{\gamma\Omega-K
p}{\gamma\Omega+K p}\right]\Bigg\}dp\nonumber,
\end{eqnarray} where $\gamma_0=(1-\Omega^2/K^2)^{-1/2}$.
The Landau rule is applied, so that $\sigma=0$ for ${\rm Re}[\Omega^2]\ge K^2$ and for ${\rm
Re}[\Omega^2]<K^2$ one has\citep{bergman}
\begin{equation}\label{landau}
\sigma=\begin{cases}
0 & \text{for ${\rm Im}[\Omega]>0,$}\\
2 & \text{for ${\rm Im}[\Omega]<0.$}
\end{cases}
\end{equation}
The case with ${\rm Im}[\Omega]=0$ for ${\rm Re}[\Omega^2]<K^2$ is unphysical\citep{landau10}.
Only one sort of species is considered in the above formulas without the loss of generality. The
plasma density $n$ is that of the mobile species in the the case of immobile ions or the total
density for electron-positron plasma.



\section{Numeric results}\label{sec_numeric}
The dispersion laws $\Omega(K)$ of waves propagating in plasma are determined as solutions of the
equations
\begin{equation}\label{dispersion}
\varepsilon_L=0, \qquad \varepsilon_T=\frac{K^2}{\Omega^2}.
\end{equation}

\subsection{Thermal distribution}
The thermal relativistic distribution of particles
\begin{equation}\label{distth}
f_T(p)=\frac{\rho\exp (-\rho\gamma)}{4\pi K_2(\rho)}
\end{equation} describes highly collisional relaxed plasma. Here and below $K_n$ represents $n$-th modified Bessel function of the second
kind.

\begin{figure}[h]
\includegraphics[width=86mm]{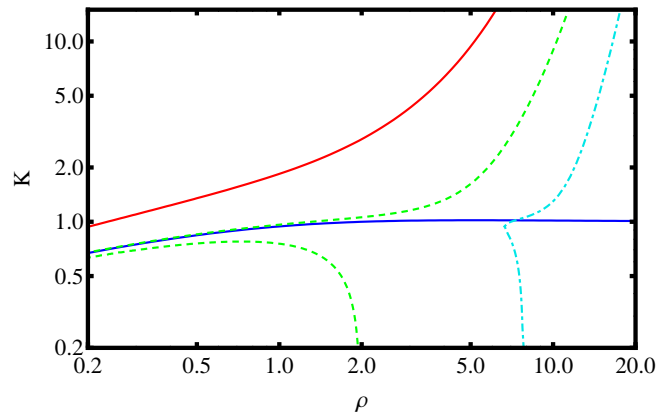}
 \caption{\label{fig:critpowerTH}(Color online) Upper solid (red) - maximum $K$ for principal damped longitudinal
 harmonic, lower solid (blue) - for undamped harmonic; dashed (green) - $K$ boundaries for second damped harmonic,
dot-dashed (cyan) - for third damped harmonic.}
\end{figure}

\begin{figure}[h]
\includegraphics[width=86mm]{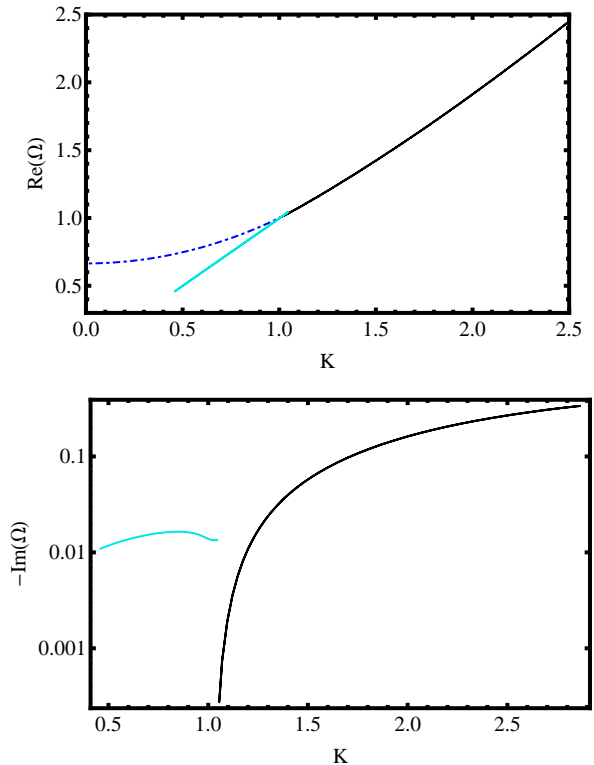}
 \caption{\label{fig:thermal2}(Color online) Dispersion relation for longitudinal waves for $\rho=2.$ Undamped mode is dot-dashed (blue), principal damped mode is
 dark solid (black), second damped mode is light solid (cyan).}
\end{figure}

The longitudinal waves in thermal plasma have variety of features. The non-relativistic
theory\citep{jackson} predicts the infinite number of damped modes, whereas the relativistic
analysis\citep{schlick_kin} limits the damped modes to a finite set. In addition, the superluminal
undamped part of the spectrum appears. The boundary $K$ for these modes are shown on
Fig.~\ref{fig:critpowerTH}. These are the solutions of the dispersion relation (\ref{dispersion})
for real $\Omega$ and $K.$ The transition between the damped (high $K$) and undamped (low $K$)
modes is indicated by the lower solid blue line. The principal damped mode exists for $K$ limited
by the upper solid red line, whose $K$ grows exponentially fast as $T\rightarrow0.$ No modes
formally exist above that line, as the ion sound is absent. However, the calculations in this
large-$K$ area are unphysical, because the set of particles constituting plasma does not behave
coherently at very low wavelengths, as the wavelength approaches the particle effective mean free
path about Debye radius $\lambda_D.$ Thus the branch ceases to exist at wavenumbers
$k\gtrsim1/\lambda_D$\citep{melrose_dispersive}. Every realization of the disribution function
$f(p)$ gives rise to different evolution of the imposed initial condition. The dashed green lines
indicate the upper and lower liming $K$ for the second damped mode. This mode exists for any
temperature. Its region of allowed $K$ overlaps with that for the principal mode. The lower
limiting $K$ goes to $0$ at $\rho\sim2$ and is zero for higher $\rho$ (lower temperature). The same
is true for the third damping mode (boundaries in dot-dashed cyan): the lower limiting $K$ goes to
$0$ at $\rho\sim8.$ The third damped mode exists only for $\rho>6.7.$

The mode completion effect was claimed to exist by Schlikeiser\citep{schlick_kin}. It consists of
the smooth transition of ${\rm Re}(\Omega(K))$ between the principal and second damped modes. This
is quite surprising given the fact that ${\rm Im}(\Omega(K))$ is discontinuous, thus $\Omega(K)$ is
discontinuous. The careful analysis shows that the real part is also discontinuous. The real and
imaginary parts of $\Omega(K)$ for $\rho=2$ are shown on Fig.~\ref{fig:thermal2} (see Fig. 3 in
Ref.~(\onlinecite{schlick_kin}) for comparison). In agreement with $K$-overlapping of the principal
and second damped modes (Fig.~\ref{fig:critpowerTH}), $K_{\rm max}\approx1.060$ for the second
damped modes is larger than $K_{\rm min}\approx1.000$ for the principal damped mode. Thus, the mode
completion effect does not occur.

\begin{figure}[h]
\includegraphics[width=86mm]{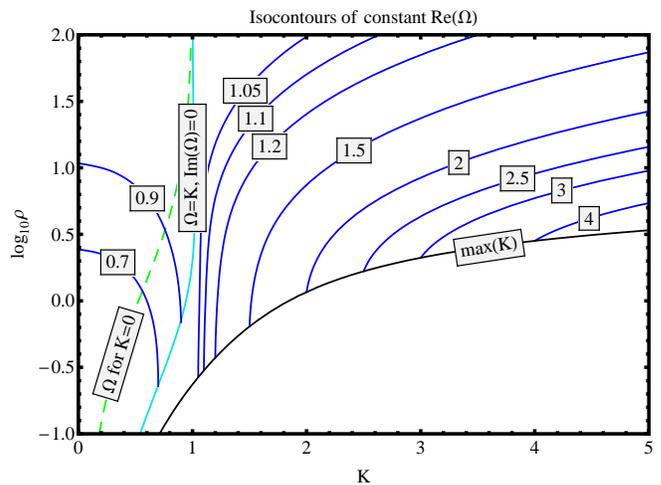}
 \caption{\label{fig:replotth}(Color online) Dispersion contours for longitudinal waves, showing ${\rm Re}(\Omega)$ in $(K, \log_{10}\rho)$ plane (solid blue lines),
 zero damping boundary (light solid cyan line), maximum $K$ curve (dark solid black line), $\Omega$ as a function of $\rho$ for $K=0$ (dashed green line).}
\end{figure}

\begin{figure}[h]
\includegraphics[width=86mm]{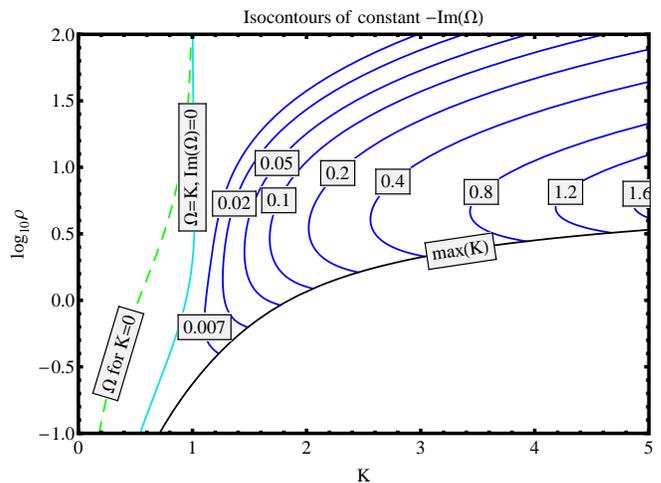}
 \caption{\label{fig:implotth}(Color online) Dispersion contours for longitudinal waves, showing $-{\rm Im}(\Omega)$ in $(K, \log_{10}\rho)$ plane (solid blue
 lines), zero damping boundary (light solid cyan line), maximum $K$ curve (dark solid black line), $\Omega$ as a function of $\rho$ for $K=0$ (dashed green line).}
\end{figure}

The isocontours of constant ${\rm Re}(\Omega)$ and $-{\rm Im}(\Omega)$ for the principal mode are
shown on, respectively, Fig.~\ref{fig:replotth} and Fig.~\ref{fig:implotth}. These are the
extensions of Figs. 6 and 7 in Ref.~(\onlinecite{bergman}) to higher temperatures (lower $\rho$)
with the maximum $K$ boundary (see Fig.~\ref{fig:critpowerTH}) employed. No modes formally exist
below the dark solid line denoted by "$\rm max(K)$". The modes are undamped to the left from the
light solid cyan line "$\rm \Omega=K, Im(\Omega)=0$" and damped to the right. The plasma frequency
at the infinite wavelength is shown in dashed green. The correspondent plot for the transverse
waves is given in Ref.~(\onlinecite{bergman}) along with the approximations.

\subsection{Power-law distribution}
The distribution
\begin{equation}\label{distpower_law}
f_P(p)=\frac{\Gamma(\kappa)}{\pi^{3/2}\Gamma(\kappa-\frac32)}\gamma^{-2\kappa}
\end{equation} represents a simplest $f(p)$. Also, it produces many analytic results (see \S \ref{sec_analytic}).
It is often used to interpret the results of astronomical observations of jets\citep{wardle} or any
object where shock acceleration takes place. In applications this distribution is usually applied
with the limited range of $\gamma,$ but for my calculation the entire range of $\gamma$ is taken.

\begin{figure}[h]
\includegraphics[width=86mm]{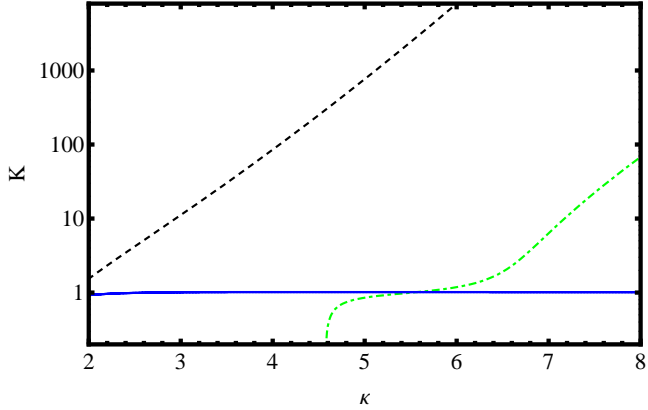}
 \caption{\label{fig:critpower}(Color online) Maximum $K$ for principal damped longitudinal mode (dashed black line),
 for second damped mode (dot-dashed green line), for undamped mode (solid blue line).}
\end{figure}

The migration of critical points of longitudinal dispersion relation is shown on
Fig.~\ref{fig:critpower}. The superluminal mode smoothly converts to a damped principal mode at
critical $K$ about $1,$ shown in solid blue on the figure. The maximum $K$ of the principal mode
(dashed black curve) grows exponentially fast as a function of $\kappa$ similarly to the behavior
of maximum $K$ boundary for the thermal distribution Fig.~\ref{fig:critpowerTH}. The behavior of
higher damped modes is, in contrast, different. The second damped mode appears at $\kappa=4.58.$ At
higher $\kappa$ second damped mode can assume $K$ from $0$ to a critical $K,$ shown in dot-dashed
green on the figure. The presence of the second and higher order damped modes at $K=0$ is a feature
of the power-law distribution and is not observed for the thermal distribution.

\begin{figure}[h]
\includegraphics[width=86mm]{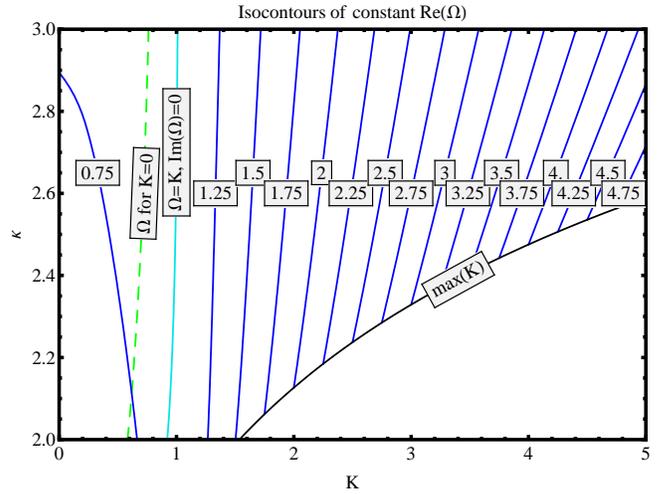}
 \caption{\label{fig:replotkk}(Color online) Dispersion contours for longitudinal waves, showing ${\rm Re}(\Omega)$ in the $(K,\kappa)$ plane (solid blue lines),
 zero damping boundary (light solid cyan line), maximum $K$ curve (dark solid black line), $\Omega$ as a function $\kappa$ for $K=0$ (dashed green line).}
\end{figure}

\begin{figure}[h]
\includegraphics[width=86mm]{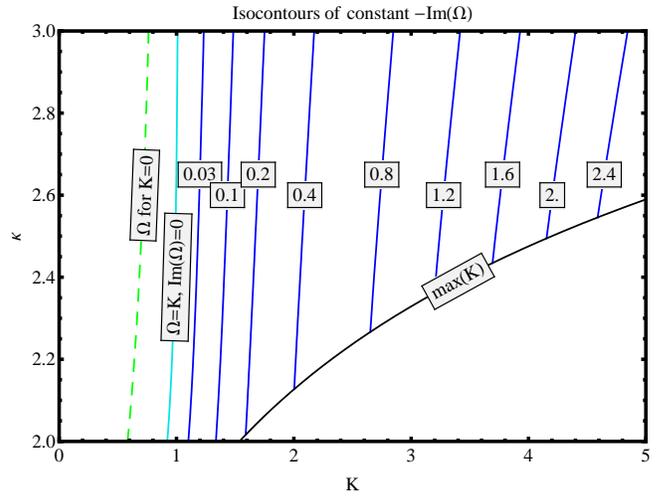}
 \caption{\label{fig:implotkk}(Color online) Dispersion contours for longitudinal waves, showing $-{\rm Im}(\Omega)$ in the $(K,\kappa)$ plane (solid blue lines),
 zero damping boundary (light solid cyan line), maximum $K$ curve (dark solid black line), $\Omega$ as a function $\kappa$ for $K=0$ (dashed green line).}
\end{figure}

The isocontours of constant ${\rm Re}(\Omega)$ and $-{\rm Im}(\Omega)$ for the principal mode are
shown on, respectively, Fig.~\ref{fig:replotkk} and Fig.~\ref{fig:implotkk}. The choosen small
range of $\kappa\in[2,3]$ reflects the astronomical need to consider distributions $N(\gamma)\sim
\gamma^{-x}d\gamma$ for $\gamma\gg1$ with $x \in [2,4].$ As in the case of thermal plasma, no modes
formally exist below the solid black line denoted by "$\rm max(K)$". The modes are undamped to the
left from the light solid cyan line "$\rm \Omega=K, Im(\Omega)=0$" and damped to the right. The
plasma frequency at the infinite wavelength is shown in dashed green. The isocontour $K$ appears to
depend only weakly on $\kappa,$ however the isocontour $-{\rm Im}(\Omega)$ values
(Fig.~\ref{fig:implotkk}) are at least $1.5$ times larger for the same $K,$ than in case of thermal
distribution (Fig.~\ref{fig:implotth}). The comparison of damping coefficients for thermal and
hybrid distributions is given in the next subsection.

\subsection{Hybrid distribution}

I choose the relativistic isotropic Lorentzian $\kappa$ distribution \cite{xiao_earth,xiao}
\begin{equation}\label{distkappa}
f_H(p)\propto\left(1+\frac{\sqrt{1+p^2}-1}{\kappa \theta^2}\right)^{-(\kappa+1)},
\end{equation} which tends to
relativistic thermal distribution with $T=\theta^2$ as $\kappa$ approaches infinity. The
normalization coefficient is calculated numerically from the condition $\int^\infty_0 4\pi p^2
f_H(p)dp=1.$ This distribution is the realistic choice when the acceleration of particles is
balanced by the relaxation processes: almost thermal distribution at high $\kappa$ and almost
power-law at low $\kappa.$ Thus, it is possible to compare the wave propagation for thermal,
slightly non-thermal and highly non-thermal distributions.

\begin{figure}[h]
\includegraphics[width=92mm]{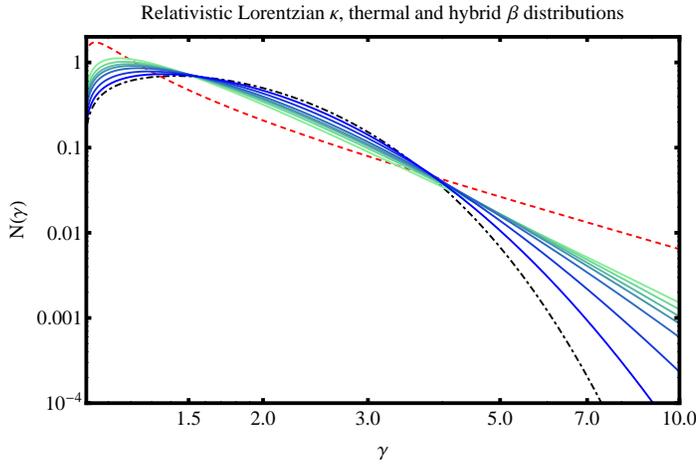}
 \caption{\label{fig:distplot}(Color online) Distributions of particles with kinetic energy $E_K=1:$ thermal ($\kappa\rightarrow\infty$) (dot-dashed black line),
Lorentzian $\kappa$ with $\kappa=4.5, 5, 5.5, 6, 7, 10, 20$ (light green to dark blue lines),
hybrid $\beta$ distribution (infinite $E_K$) (dashed red line).}
\end{figure}

The comparison of distributions is made on Fig~\ref{fig:distplot}. The inverse temperature $\rho$
for $f_T(p)$ and the temperature-like parameter $\theta^2$ for $f_H(p)$ are choosen in such a way
that the average kinetic energy of particle $E_K=\int (\gamma-1)4\pi p^2 f(p)dp$ is equal to $1.$
The choice of constant $E_K=1$ is made to show the typical behavior of dispersion relations for
plasmas with approximately the same energetics. The thermal distribution is shown in dot-dashed
black. The transition from light green to dark blue corresponds to the increase of $\kappa$ through
the discrete set $\kappa=4.5, 5, 5.5, 6, 7, 10, 20.$ The relativistic Lorentzian $\kappa$
distribution appears to have more particles at lower $\gamma\approx1.2$ and higher
$\gamma\gtrsim5,$ but fewer particles in the intermediate region $\gamma\sim 3.$ For the contrast I
added the hybrid $\beta$ distribution
\begin{equation}\label{distbeta} f_{H2}(p)=\frac{\beta^3}{\pi^2 (1+\beta^2p^2)^2}
\end{equation} with $\beta=2$ that is shown by the dashed red line. It was not normalized to $E_K=1,$ since it has the divergent total
energy. However, the dispersion laws exist for it.

\begin{figure}[h]
\includegraphics[width=92mm]{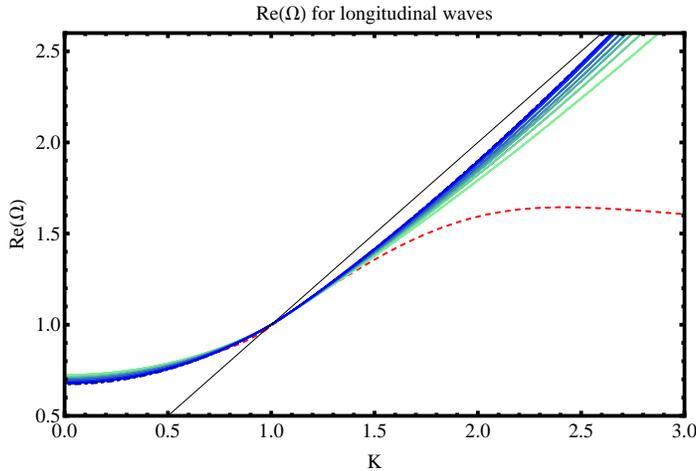}
 \caption{\label{fig:lorentzre}(Color online) ${\rm Re}(\Omega(K))$ dependence for longitudinal waves for
 Lorentzian $\kappa$ distribution: limiting thermal case ($\kappa\rightarrow\infty$) (dot-dashed black line),
  various $\kappa$ from $4.5$ to $20$ (light green to dark blue lines), hybrid $\beta$ distribution (dashed
red line).}
\end{figure}

\begin{figure}[h]
\includegraphics[width=92mm]{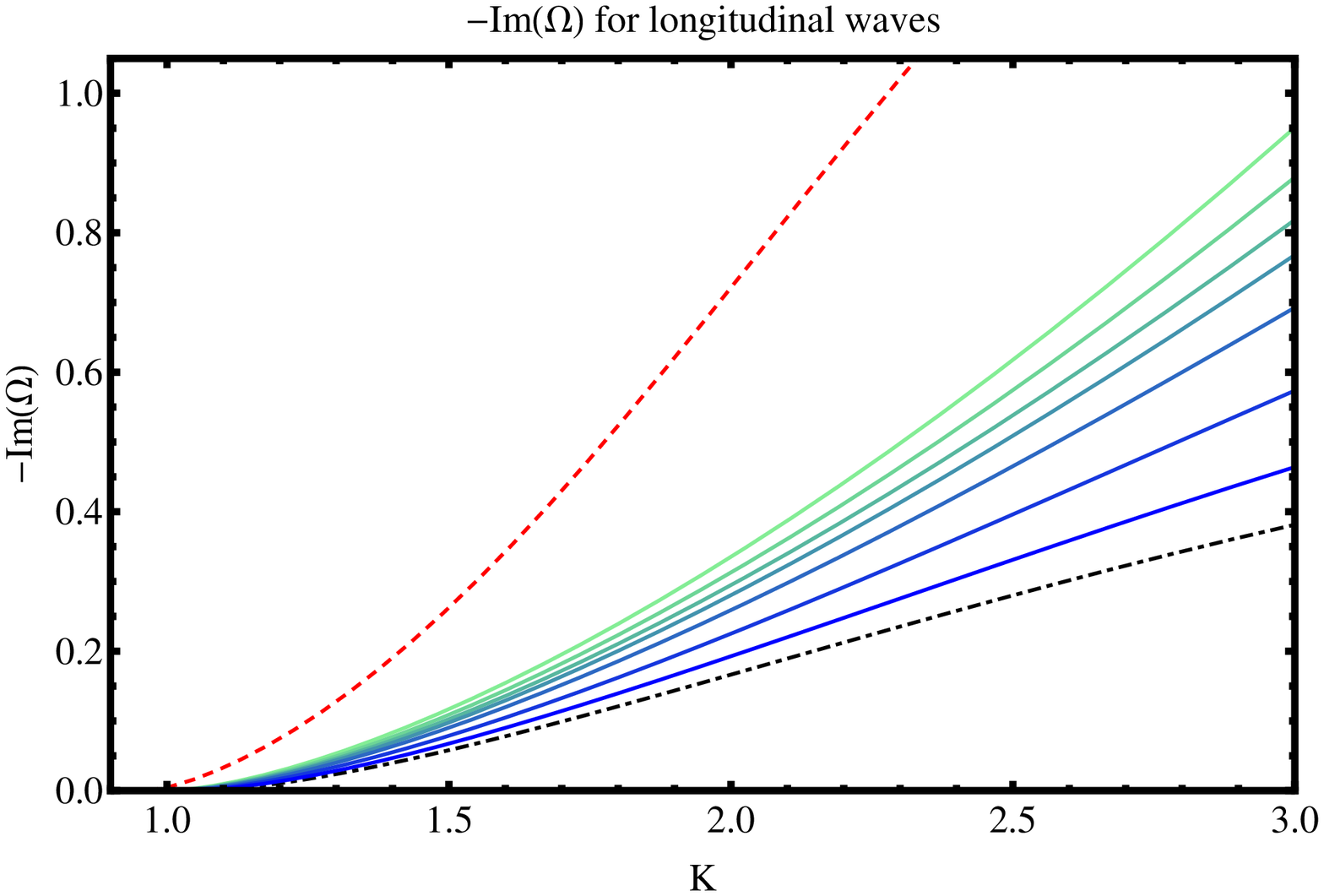}
 \caption{\label{fig:lorentzim}(Color online) ${\rm -Im}(\Omega(K))$ dependence for longitudinal waves for
 Lorentzian $\kappa$ distribution: limiting thermal case ($\kappa\rightarrow\infty$) (dot-dashed black line),
 various $\kappa$ from $4.5$ to $20$ (light green to dark blue lines), hybrid $\beta$ distribution (dashed
red line).}
\end{figure}

\begin{figure}[h]
\includegraphics[width=92mm]{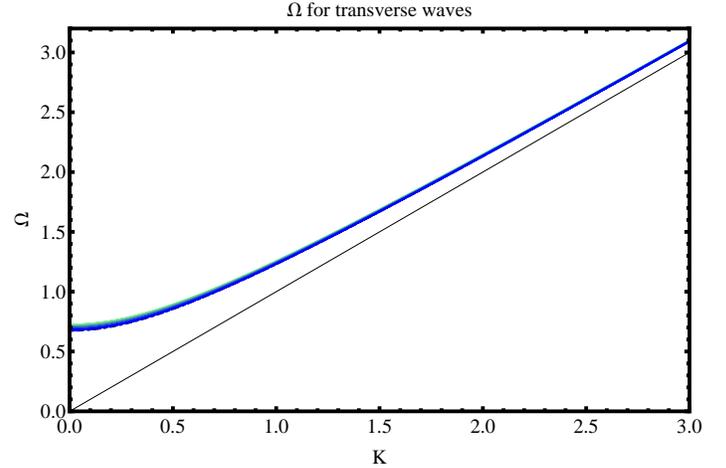}
 \caption{\label{fig:lorentztr}(Color online) ${\rm Re}(\Omega(K))$ dependence for transverse waves for
 Lorentzian $\kappa$ distribution: limiting thermal case ($\kappa\rightarrow\infty$) (dot-dashed black line),
 various $\kappa$ from $4.5$ to $20$ (light green to dark blue lines), hybrid $\beta$ distribution (dashed
 red line).}
\end{figure}
The dependencies of ${\rm Re}(\Omega(K))$ for longitudinal waves, $-{\rm Im}(\Omega(K))$ for
longitudinal waves and $\Omega(K)$ for transverse waves are shown for principal modes on,
respectively, Fig.~\ref{fig:lorentzre}, Fig.~\ref{fig:lorentzim}, Fig.~\ref{fig:lorentztr}. The
thin black line on Fig.~\ref{fig:lorentzre} separates the undamped mode (on the left) from the
damped one (on the right). The long-wavelength behavior of longitudinal modes is predictable: more
mobile lower $\gamma$ species in thermal distribution have the lowest plasma frequency. By the same
reason $\Omega$ at $K=0$ for transverse waves (Fig.~\ref{fig:lorentztr}) is larger for smaller
$\kappa.$ The variation of ${\rm Re}(\Omega)$ is within $20\%$ for both longitudinal and transverse
waves. The Landau damping (Fig.~\ref{fig:lorentzim}) shows larger variation of up to $3$ times. The
electrons responsible for Landau damping mainly have high $\gamma\gtrsim5,$ what makes
distributions with smaller $\kappa$ dissipate waves more effectively. The similar result is
observed for non-relativistic Lorentzian $\kappa$ distribution\citep{mace}. The behavior of plasma
waves for hybrid $\beta$ distribution is very different from that for thermal or Lorentzian
$\kappa$ distribution.

\section{Analytic formulas}\label{sec_analytic}
\subsection{Thermal distribution}
 The Trubnikov's form\cite{trubnikov} of the dielectric tensor
reads
\begin{equation}\label{trubnikov}
\varepsilon_{\mu\nu}=\delta_{\mu\nu}+\frac{i}{\Omega^2}\frac{\rho^2}{K_2(\rho)}\int^\infty_0
d\alpha \left(\frac{K_2(r)}{r^2}\delta_{\mu\nu}-\alpha^2\frac{K_\mu K_\nu}{\Omega^2}
\frac{K_3(r)}{r^3}\right),
\end{equation} where
\begin{equation}
r=\left[\rho^2-2i\alpha\rho+\alpha^2\left(\frac{K^2}{\Omega^2}-1\right)\right]^{1/2}.
\end{equation}
I denote
\begin{equation}\Delta=1-\frac{K^2}{\Omega^2}\end{equation} and expand the Bessel functions in equation (\ref{trubnikov})
in series in $\Delta.$ The resultant integrals can be evaluated analytically knowing that
\begin{equation}
\int^\infty_0 d\alpha\left[ \alpha^n
\frac{K_m(\sqrt{\rho^2-2i\rho\alpha})}{(\rho^2-2i\rho\alpha)^{m/2}}\right]=n!
i^{n+1}\frac{K_{n-m+1}(\rho)}{\rho^m}.
\end{equation}

Then the implicit formula for $\Delta$
\begin{equation}
\Delta=\Omega^{-2}\sum^\infty_{j=0}\left(-\frac{\Delta}{2\rho}\right)^j\frac{(2j)!}{j!}\frac{K_{j-1}(\rho)}{K_2(\rho)}
\end{equation} can be derived \cite{imre}.
 It can be rewritten as the explicit formula for $\Delta.$ The expression for the transverse permittivity
\begin{equation}
\varepsilon_T=1-\Delta=1-\sum^\infty_{n=0}\Omega^{-2(n+1)}\prod^n_{l=0}\frac{K_{l-1}(\rho)}{K_2(\rho)}\frac{(2l)!}{(-2\rho)^ll!}
\end{equation} can easily give the dependence $K(\Omega).$ Taking $n=0$ one obtains
\begin{equation}\label{nonmagn}
\Omega^2=K^2+ \frac{K_1(T^{-1})}{K_2(T^{-1})}
\end{equation} in a high-frequency limit $\Omega\gg 1$, coincident with the expression in
Ref.~(\onlinecite{melrosec_q}). As the frequency goes down, terms with higher $n$ become important.
The consecutive approximations are expected to work well for any temperature. The author believes
this elegant derivation of the high-frequency approximation was not previously outlined.
\subsection{Power-law distribution} The permittivity tensor for the power-law distribution (\ref{distpower_law}) of particles can be calculated
analytically.


\subsubsection{General case}
 The permittivity components are

\begin{subequations}
\begin{eqnarray}
\varepsilon_L&=&1+\frac{2\Gamma(\kappa-1)}{\Gamma(\kappa+\frac12)\Gamma(\kappa-\frac32)K^2}\bigg\{\Gamma(\kappa+1)\\
&-&\kappa\csc(\pi\kappa)\bigg[\sqrt{\pi}(1-\Delta)^{\frac12-\kappa}\Delta^{\kappa-1}\Gamma\bigg(\kappa+\frac12\bigg)\nonumber\\
&+&\pi\bigg(\kappa-\frac12\bigg)\nonumber \bigg(2{}_2\tilde{F}_1\bigg(2,-\frac12,2-\kappa,\Delta\bigg)\nonumber\\
&-&3{}_2\tilde{F}_1\bigg(1,-\frac12,2-\kappa,\Delta\bigg)\bigg)\bigg]\nonumber\\
&+&\sqrt{\pi}\Delta^{\kappa-1}(\Delta-1)^{\frac12-\kappa}\kappa\sigma\Gamma\bigg(\kappa+\frac12\bigg)
\bigg\}\nonumber,
\end{eqnarray}
\begin{eqnarray}
\varepsilon_T&=&1-\frac{\Gamma(\kappa-1)}{2\Gamma(\kappa+\frac12)\Gamma(\kappa-\frac32)K^2}\bigg\{2\Gamma(\kappa+1)\\
&-&\csc(\pi\kappa)\bigg[2\sqrt{\pi}\Delta^\kappa(1-\Delta)^{\frac12-\kappa}\Gamma\bigg(\kappa+\frac12\bigg)\nonumber\\
&+&\pi\kappa(2-3\Delta-2\kappa){}_2\tilde{F}_1\bigg(1,-\frac12,2-\kappa,\Delta\bigg)\nonumber\\
&+&2\pi(\Delta\kappa+\kappa-1){}_2\tilde{F}_1\bigg(2,-\frac12,2-\kappa,\Delta\bigg)
\bigg]\nonumber\\
&-&2\sqrt{\pi}\Delta^\kappa(\Delta-1)^{\frac12-\kappa}\sigma\Gamma\bigg(\kappa+\frac12\bigg)\bigg\}\nonumber,
\end{eqnarray}
\end{subequations} where $\sigma$ is determined by Landau rule (see equation (\ref{landau}) and discussion therein).
The notation ${}_2\tilde{F}_1$ represents the regularized hypergeometric function. For the integer
$\kappa$ the limit of $\kappa$ going to that singular integer value should be considered. The
non-singular expressions were also derived, but are longer and are not provided here. The
Mathematica 6 convention of branch cuts should be used to evaluate the values of roots and
non-integer powers. It sets the branch cuts to be on ${\rm arg}(z)=\pi$ line in the complex z
plane.

The high-frequency limit for the transverse waves is
\begin{equation}
\Omega^2=K^2-\frac{\pi\csc(\pi\kappa)\Gamma(\kappa)}{\Gamma(2-\kappa)\Gamma(\kappa-\frac32)\Gamma(\kappa+\frac12)}.
\end{equation} Again the limit must be considered for integer $\kappa.$

\subsubsection{Special cases}
Much shorter formulas for permittivities can be derived for certain $\kappa.$ The shortest ones are
for $\kappa$ in the observationally motivated range $\kappa \in [2,3].$ Let me choose $\kappa=2$
and $\kappa=5/2$ as the examples.

The case $\kappa=2$ (equivalent to $dN(\gamma)\sim \gamma^{-2}d\gamma$ at high $\gamma$) leads to
\begin{subequations}
\begin{equation}
\varepsilon_T=1+\frac{(4-10\Delta)\sqrt{\Delta-1}+3\sigma\pi\Delta^2+6\Delta^2{\rm
arcsec}(\sqrt{\Delta})}{3\pi K^2(\Delta-1)^{3/2}},
\end{equation}
\begin{equation}
\varepsilon_L=1+\frac{(8+16\Delta)\sqrt{\Delta-1}+12\sigma\pi\Delta^2-24\Delta{\rm
arcsec}(\sqrt{\Delta})}{3\pi K^2(\Delta-1)^{3/2}}.
\end{equation}
\end{subequations} The case $\kappa=\frac52$ (equivalent to $dN(\gamma)\sim \gamma^{-3}d\gamma$ at high $\gamma$) leads to
\begin{subequations}
\begin{equation}
\varepsilon_T=1+\pi\frac{\Delta(10-15\Delta+8\Delta^{3/2}(1-2\sigma))-3}{16K^2(\Delta-1)^2},
\end{equation}
\begin{equation}\label{long52}
\varepsilon_L=1+5\pi\frac{\Delta(6+3\Delta-8\Delta^{1/2}(1-2\sigma))-1}{16K^2(\Delta-1)^2}.
\end{equation}
\end{subequations}

\begin{figure}[h]
\includegraphics[width=92mm]{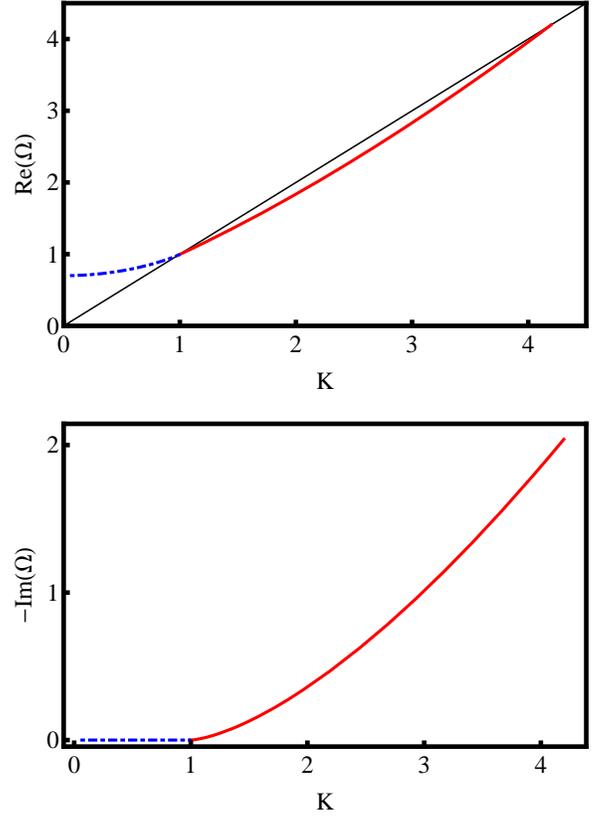}
 \caption{\label{fig:longit52}(Color online) Longitudinal dispersion law for $\kappa=5/2$: undamped mode (dot-dashed blue line),
 damped mode (solid red line), separated by ${\rm Re}(\Omega)=K$ (thin black line).}
\end{figure}

The longitudinal dispersion relation (\ref{long52}) for $\kappa=5/2$ remarkably gives the compact
analytic form for $\Omega(K).$ The undamped mode at low $K$ exists in this case along with the
single damped mode at higher $K$ as
\begin{subequations}\label{longres52}
\begin{eqnarray}
\Omega=\frac18\sqrt{48K^2+5\pi+\sqrt{\frac{(5\pi-16K^2)^3}{5\pi}}}\\
\text{for}\quad K=0..\frac{\sqrt{5\pi}}4\nonumber,
\end{eqnarray}
\begin{eqnarray}
\Omega=\frac18\sqrt{48K^2+5\pi-i\sqrt{\frac{(16K^2-5\pi)^3}{5\pi}}}\\
\text{for}\quad K=\frac{\sqrt{5\pi}}4.. \frac14\sqrt{5\pi(9+4\sqrt{5})}.\nonumber
\end{eqnarray}
\end{subequations} The relation $\Omega(K)$ can be expanded near the point $K_x=\sqrt{5\pi}/4$ to
give the powers of $(K_x-K)^{3/2}$ so that $\Omega_K$ stays real for $K<K_x$ and gains the
imaginary part for $K>K_x.$ The second derivative of $\Omega(K)$ is discontinuous at $K_x.$ The
explicit dispersion law $\Omega(K)$ (\ref{longres52}) is plotted for $\kappa=5/2$ on
Fig.~\ref{fig:longit52}.

\subsection{Hybrid $\beta$ distribution}
The analytic expressions for transverse and longitudinal permittivities need some terms to be
abbreviated for brevity in the case of hybrid $\beta$ distribution (\ref{distbeta}). They read
\begin{subequations}
\begin{eqnarray}
\varepsilon_T&=&1+\frac\beta{\pi A^3B C K^2}\bigg[C(2A B+\beta^4((B-2C^2)\\
&\times& \arctan A -B {\rm arcsec}\beta))-2A^3\Delta^2{\rm arctanh}
C\bigg]\nonumber\\&-&\frac{\Delta^2\beta\sigma i}{B C K^2}\nonumber,
\end{eqnarray}
\begin{eqnarray}
\varepsilon_L&=&1+\frac{2\beta}{\pi A^3B^2C K^2}\bigg[B^2C\beta^4 {\rm arcsec}\beta-A C\\
&\times&(B(B+\Delta A^2)+A\beta^4(B+2C^2)\arctan A)\nonumber\\
&+&2A^3\Delta(B+C^2\beta^2){\rm arctanh}C\bigg]\nonumber\\
&-&\frac{2\beta\Delta\sigma i(B+C^2\beta^2)}{B^2 C K^2}\nonumber,
\end{eqnarray}
where
\begin{equation}
A=\sqrt{\beta^2-1}, \quad B=\beta^2-\Delta, \quad C=\sqrt{1-\Delta}. \end{equation}
\end{subequations}
The expressions for $\beta=1$ coincide with those for the power-law distribution with $\kappa=2.$
\section{Discussion \& Conclusion}\label{sec_discussion}
The paper derives and reconsiders the broad range of linear wave effects in non-magnetized one
species plasma. The various distributions are employed for comparison and to consider the realistic
non-equilibrium plasmas. The thermal (\ref{distth}), power-law (\ref{distpower_law}), relativistic
Lorentzian $\kappa$ (\ref{distkappa}) and hybrid $\beta$ (\ref{distbeta}) distributions are
employed with the possible inclusion of the simulated distributions\cite{liu} in the future.

For longitudinal waves the maximum $K$ was found to exist above that neither the undamped mode nor
the damped mode survive. The mode completion effect\citep{schlick_kin} was closed. The regions of
the principal and the second damped modes overlap in the temperature-wavenumber plane for thermal
plasma. The second damped mode always exist. The lower possible $K$ is zero for low temperatures
for second and third damped modes, but is non-zero for higher temperatures. In contrast, for the
power-law distribution (\ref{distpower_law}) the second damped mode does not exist for low
$\kappa$. When it exists at higher $\kappa$, its lowest $K$ is always zero.

Some analytic results are derived that can accelerate the calculations and provide some insight.
The full analytic result for longitudinal $\Omega(K)$ was derived for the power-law distribution
(\ref{distpower_law}) with $\kappa=5/2.$ The transition at $K_x$ between the undamped mode and the
principle damped mode was found to be smooth with discontinuous second derivative of $\Omega(K)$
because of $(K_x-K)^{3/2}$ term.

 \acknowledgements The author is grateful to Ramesh Narayan for fruitful discussions and to
 anonymous referee, whose comments helped to improve the manuscript. The paper was supported by NASA Earth
 and Space Science Fellowship NNX08AX04H and partially supported by NASA grant NNX08AH32G.

\end{document}